\documentstyle[12pt,aaspptwo,psfig]{article}
\def\etal{{\it et al. }}

\begin{document}

\title{Hubble Space Telescope Imaging of the Globular Cluster 
System around NGC 5846\altaffilmark{1}}

\author{Duncan A. Forbes}
\affil{Lick Observatory, University of California, Santa Cruz, CA 95064}
\and
\affil{School of Physics and Space Research, University of Birmingham,
Edgbaston, Birmingham B15 2TT, United Kingdom} 
\affil{Electronic mail: forbes@lick.ucsc.edu}

\author{Jean P. Brodie}
\affil{Lick Observatory, University of California, Santa Cruz, CA 95064}
\affil{Electronic mail: brodie@lick.ucsc.edu}

\author{John Huchra}
\affil{Harvard--Smithsonian Center for Astrophysics, 60 Garden Street,
Cambridge, Massachusetts 02138}
\affil{Electronic mail: huchra@cfa.harvard.edu}

\altaffiltext{1}{Based on observations with the NASA/ESA {\it Hubble
Space Telescope}, obtained at the Space Telescope Science Institute,
which is operated by AURA, Inc., under NASA contract NAS 5--26555}

\begin{abstract}

Bimodal globular cluster metallicity distributions have now been seen
in a handful of large ellipticals. Here we report the discovery of a 
bimodal distribution in the dominant group elliptical NGC 5846, using 
the {\it Hubble Space Telescope's} Wide Field and Planetary
Camera 2 (WFPC2). The two peaks are located at V--I = 0.96 and 1.17,
which roughly correspond to metallicities of [Fe/H] = --1.2 and --0.2
respectively. The luminosity functions of the blue and red
subpopulations appear to be the same, indicating that luminosity does
not correlate with metallicity within an individual 
galaxy's globular cluster system. 

Our WFPC2 data cover three pointings allowing us to
examine the spatial distribution of globular clusters out to 30 kpc
(or 2.5 galaxy effective radii). 
We find a power law surface density with a very flat slope, and a
tendency for globular clusters to align close to the galaxy minor
axis. An extrapolation of the surface density profile, out to 50 kpc,
gives a specific frequency S$_N$ = 4.3 $\pm$ 1.1. 
Thus NGC 5846 has a much lower
specific frequency than other dominant ellipticals in clusters but is
similar to those in groups. 
The central galaxy regions reveal some filamentary dust features, 
presumably from a past merger or accretion of a gas--rich galaxy. 
This dust reaches to the very nucleus and so provides an obvious
source of fuel for the radio core. We have searched for
proto--globular clusters that may have resulted from the
merger/accretion and find none. 
Finally, we briefly discuss the implications of our results for 
globular cluster formation mechanisms. 

\end{abstract}

\section{Introduction}

One of the most interesting recent developments in globular cluster 
(GC) studies is the discovery that at least some elliptical galaxies 
have a bimodal 
GC metallicity distribution. A bimodal distribution is also seen in 
the Milky Way GC system. 
These bimodal, and sometimes possibly trimodal, distributions are 
more metal--rich 
in the mean than those of the Milky Way but the same general 
conclusion can be drawn from their existence;  
galaxies with multimodal metallicity distributions 
have not undergone a simple monolithic 
collapse but have experienced episodic star 
formation. To date, the best examples of multimodal GC metallicity 
distributions are cD galaxies with extremely rich GC 
systems (i.e. with specific frequency S$_N$ $\sim$ 15) such as M87 in 
Virgo (McLaughlin, Harris \& Hanes 1994), NGC 1399 in Fornax 
(Bridges, Hanes \& Harris 1991) and NGC 3311 in 
Hydra (McLaughlin \etal 1995). 
Two galaxies with `normal' specific frequencies 
have also revealed bimodal distributions i.e., NGC 4472 with S$_N$ =
5.6 (Geisler \etal 1996) 
and NGC 3923 with S$_N$ = 6.4 (Zepf, Ashman \& Geisler 1995). 
Thus the formation mechanism responsible for two 
distinct GC populations must be able to explain both the populous GC
systems in cD galaxies, and the `normal' richness GC systems 
in ellipticals. 

An interesting galaxy in this regard is NGC 5846. Although it
dominates a small, compact group of galaxies it has an average or
slightly below average specific frequency. 
Detection of a bimodal GC color distribution in this galaxy 
would add a further
constraint to GC formation mechanisms. The galaxy is not a cD but a
giant E0 elliptical, with an 
effective radius of 11.6 kpc (Bender, Burstein \& Faber 1992) 
and an absolute magnitude of
M$_V$ = --22.6 (Faber \etal 1989), assuming a 
distance modulus (m--M) = 32.3 from Forbes, Brodie \& Huchra (1996;
hereafter Paper I). This would place NGC 5846 some 13 Mpc more distant
than the Virgo cluster. 
Harris \& van den Bergh (1981)
estimated the total GC count, from a photographic study, 
to be 2200 $\pm$ 1300. For M$_V$ =
--22.6 this corresponds to 
S$_N$ = 2.0 $\pm$ 1.2. (Note that Harris \& van den Bergh used M$_V$ =
--21.7 and hence derived S$_N$ = 4.5 $\pm$ 2.6.)  

The inner region of NGC 5846 contains some dust and
a compact radio core
with a possible extended component (Mollenhoff, Hummel \& Bender
1992). 
The X--ray emission from hot gas extends 
$\sim$ 100 kpc from the center of the galaxy, 
encompassing two other early--type galaxies in the group 
(Biermann, Kronberg \& Schumutzler 1989). 
Biermann \etal calculated the hot gas mass to be
$\sim$ 10$^{11}$M$_{\odot}$, and the total virial mass to be 
almost 10$^{13}$M$_{\odot}$. Recent ROSAT mapping reveals that the
soft X--ray emission is spatially coincident with the H$\alpha$
emission and dust
morphology (Goudfrooij \& Trinchieri 1996). 
The galaxy shows many characteristics similar to those of M87, which
lies at the
center of the Virgo cluster. Both are luminous galaxies (M$_V$ =
--22.5 for M87) which dominate the local potential and lie at the
center of an extensive X--ray envelope. The virial mass and cooling
flow infall rate are also similar for the two galaxies. 
The central velocity dispersion is 361 km s$^{-1}$ for M87 and 278 km
s$^{-1}$ for NGC 5846 (Faber \etal 1989). 
They differ greatly in one important regard, namely GC specific
frequency (S$_N$ = 14 $\pm$ 0.5 for M87). 
It has been
speculated that some of M87's GCs were formed as a result of its
location at the center of a large gravitational potential, either from
the cooling flow (Fabian, Nulsen \& Canizares 1984), or by 
accretion from neighboring galaxies
(e.g. Forte \etal 1982) or from the field (West \etal 1996). 
Knowledge of the GC system properties of NGC 5846, 
particularly in comparison with giant ellipticals in 
different environments, may provide
important clues about globular cluster and galaxy formation histories.

In paper I we presented the V band WFPC2 data for NGC 5846. 
The GC luminosity function (GCLF) was found to be the same in each of 
three separate pointings, ranging from $\sim$ 3 to 30 kpc from the 
galaxy center. While taking into account incompleteness, photometric errors
and background contamination we fit the combined GCLF using the maximum 
likelihood technique of Secker \& Harris (1993). This gave a turnover 
magnitude of $m^0_V$ = 25.05 $\pm$ 0.10 and a dispersion of $\sigma$ =
1.34 $\pm$ 0.06. 
Here we include the WFPC2 I band 
data and discuss the color and spatial distributions of the GC system in 
NGC 5846. 

\section{Observations and Data Reduction}

Details of the observations and data reduction are described in Paper I. 
Briefly, we obtained WFPC2 images of NGC 5846 at three pointings, in two 
filters. The pointings are located $\sim$ 1.5 arcmin north of the nucleus, 
$\sim$ 2.5 arcmin south and one with the galaxy centered in the PC CCD. For
each filter, we have a pair of images giving a total of 2400 seconds 
in F555W and 2300 seconds 
in F814W. 
Objects were detected using 
DAOPHOT (Stetson 1987) with a 
S/N threshold of 4 per pixel. 
Aperture magnitudes (in a 2 pixel radius) were converted to total 
Johnson V and Cousins I mags. A Galactic 
extinction correction (A$_V$ = 0.11 and A$_I$ = 0.05) was also applied. The 
DAOPHOT--selected object list was checked for matches to hotpixels and 
all sources with FWHM greater than 3 pixels were removed. 
Contamination in the final object list is expected to be less than 5\%. 
Completeness tests for the V band data indicate a 50\% incompleteness at
V = 25.95. Photometric errors at this magnitude are 0.16. 
In this paper we include 
GCs in the central PC which are located at galactocentric distances of 
greater than 7$^{''}$ (1 kpc) i.e., beyond the strongest dust lanes. 
Globular clusters 
located interior to this radius are discussed separately in section
4.1.

\section{Results}

\subsection{The Galaxy}

We have subtracted a smooth 
model of the galaxy from the PC CCD image of the central pointing.  
A grey scale of the residuals is shown in Fig. 1
(Plate **) which reveals several GCs along with filamentary dust
features. The DAOPHOT selected GCs, beyond 7$^{''}$ of the galaxy
center, are indicated. In Fig. 2 we give radial profiles of some
parameters from the I band galaxy model (which is less effected by dust than
the V band). The surface brightness profiles show that the V--I color
is roughly constant between 0.1 and 3 kpc. Both V and I profiles show
a change in slope at about 0.3 kpc or 2.3$^{''}$. The ellipticity
profile confirms that the galaxy is E0--1 in the central regions. 
The position angle changes from
70$^{\circ}$ to $\sim$ 100$^{\circ}$ near the nucleus, 
and 4th cosine term (see for example Forbes \& Thomson 1992) 
becomes more boxy which indicates 
extra light at 45$^{\circ}$ to the major axis.

\subsection{The Globular Clusters}

The color--magnitude diagram for our full sample of 837 GCs, from all 
three pointings, is shown in Fig. 3. The GCs range from V $\sim$ 21 to 
V $\sim$ 26. There is a hint of bimodality, best seen between 
24 $<$ V $<$ 25, otherwise the average color is 
relatively constant with magnitude. 

The mean color, weighted by photometric error, is V--I = 1.15 
with a rms scatter of $\pm$ 0.2 $\pm$ or in terms of an error on the
mean of $\pm$ 0.01. In the subsequent analysis, we have confined 
the sample to have colors within 3$\sigma$ of the mean. This excludes 63 
objects with extreme blue and red colors. These objects appear to be 
distributed randomly throughout the three pointings, in particular they are 
{\it not} centrally located. The objects with the extreme colors 
tend to be the 
faintest sources, suggesting that photometric errors have affected their 
colors. 
Histograms for the color--selected subsample are shown in Fig. 4.
Here a color histogram is given for each of the pointings and 
for all three pointings combined. A clear bimodal color distribution is 
seen in the north and south pointings (and is reflected in the combined 
histogram), with peaks located at V--I = 0.96 and 1.17. 
The central pointing reveals the red peak, with a hint of some
additional blue GCs. 

Bearing in mind that the conversion of color into metallicity is
uncertain, we use the relation of Couture \etal (1990) to derive 
[Fe/H] from V--I. The rms error in this relation ($\sim$ 0.25 dex) 
dominates over the
error associated with measuring the mean GC color. 
Their relation gives a mean metallicity for the GC system of 
[Fe/H] = --0.3 $\pm$ 0.25. In Fig. 5 we show GC 
metallicity as a function of galactocentric distance expressed in kpc.
A weighted fit to [Fe/H] versus log of the distance in arcsec gives a
linear slope of --0.13 $\pm$ 0.03 
over the radial range of 1 to 30 kpc. This indicates that the gradient
is significant at the 4$\sigma$ level.

The surface density distribution of GCs around NGC 5846 has been
calculated using two different methods. The first, applied to the
central pointing, is to make use of the fact that out to some radius
the WFPC2 gives almost complete 180$^{\circ}$ coverage for one
hemisphere of the galaxy. This hemisphere is $\pm$ 90$^{\circ}$ about
the WFPC2 V3 axis. Thus we have counted up the GCs in six half--annuli from
1 to 13 kpc. The number in each half--annulus is doubled to give the full
360$^{\circ}$ coverage, and divided by the area of the annulus. For
the two offset pointings, the orientation of the WFPC2 field-of-view
with respect to the galaxy is much more complicated. In the second
method, we simply count up the number of GCs per CCD and divide by the
CCD area. As the GCs are not uniformly distributed across the CCD, we
use the median galactocentric distance as a weighted distance. Surface
densities from the two methods are given in Fig. 6 which shows a
smooth transition between the two methods. 
A fit to the surface density distribution using\\

$\rho = \rho_o r^{\alpha}$\\

\noindent
gives $\alpha$ = --0.7 $\pm$ 0.1 and $\rho_o$ = 4.8 $\pm$ 0.5
(assuming negligible background contamination). 

We can use the surface density profile to estimate
the total number of GCs around NGC 5846. First we must correct for the
faint undetected GCs. From our GCLF fitting in Paper I we
estimate that $\sim$ 25\% of GCs were not detected. Thus the normalization
constant $\rho_o$ should be increased to 6.0. The uncertainty in
calculating the total number of GCs is dominated by the choice 
of the outer radius. The GC counts of Harris \& van den Bergh (1981) 
reached the
background level at $\sim$ 6$^{'}$ (50 kpc). Integrating the profile
out to this radius gives a total of 4670 GCs and a corresponding 
specific frequency of S$_N$ = 4.3
$\pm$ 1.1. Integrating to 
60 kpc would increase the number of GCs by 1270. 
In addition to a global S$_N$, we can also estimate S$_N$ as a
function of radius from our central pointing. Using the integrated
galaxy V magnitude at various radii, from the catalog of Longo \& de
Vaucouleurs (1983), we can calculate the `local' S$_N$ value from our
GC counts. The S$_N$ values as a function of galactocentric distance
are shown in Fig. 7.  

The radial distribution of V magnitude is shown in Fig. 8. 
The mean magnitude shows little change with distance from the galaxy
center. The azimuthal distribution of GC counts is shown in Fig. 9. 
Here we have restricted GCs to lie within 13 kpc and between position
angles of --25$^{\circ}$ and 155$^{\circ}$ in order to obtain
complete coverage in one hemisphere. We find a slight enhancement
of GCs around 20$^{\circ}$ and a deficit around 100$^{\circ}$. 

\section{Discussion}

\subsection{Central Regions}

The galaxy central regions reveal a number of GCs and 
filamentary dust. 
The central dust indicates a past merger 
or accretion of a gas--rich galaxy and appears to reach right
into the nucleus. Given the short dynamical timescale, for galactocentric
distances of $<$ 0.5 kpc, the dust (and gas) provides an obvious
source of fuel for the compact radio core (Mollenhoff \etal 1992). 
A number of 
proto--globular clusters have been detected in currently merging galaxies
(e.g. Holtzman \etal 1992;  Whitmore \etal 1993; Whitmore \& Schweizer 
1995; Holtzman \etal 1996), as predicted by the Ashman \& Zepf (1992)
merger model. 
These objects, initially blue and very bright, 
will fade and redden with time. Thus depending on the epoch of
formation of GCs, 
we may expect to see a population of very blue or very red GCs. 
A search for proto--globular clusters in NGC 5846 is problematic due
to the dust. 
As mentioned in section 2, we excluded regions 
within 7$^{''}$ (1 kpc) of the galaxy nucleus from our object list. 
We have decided to visually inspect the inner 7$^{''}$ in {\it both} the V and 
I images for GCs. Although not objective, the eye is good at recognizing 
the brighter sources. We find 9 objects, some on the edges of dust lanes.
The V magnitudes and colors range from 23 $<$ V $<$ 24 and 0.9 $<$ V--I $<$ 
1.7. The mean color is V--I = 1.15 which is the same as the mean for
the full sample. 
Thus we find {\it no} evidence for a centrally concentrated population of GCs 
with extreme colors. Furthermore, if we assume that all objects are
instrinsically redder than V--I = 0.9, then we can place a lower limit
on their age from stellar population models. Using Bruzual \& Charlot
(1996), a cluster with V--I $>$ 0.9 and solar metallicity is at least
1 Gyr old. This suggests that {\it if} these central GCs formed from a
gas--rich merger, then it occurred over 1 Gyr ago. The spectroscopic
study 
of NGC 5846 by Fisher, Franx \& Illingworth (1995) would tend
to support this idea, as they found no evidence for a young stellar
population and concluded that the stars were uniformly old ($\sim$ 15
Gyr).

The galaxy surface brightness profiles (Fig. 2) 
do {\it not} continue to
rise with the same slope in the very inner regions i.e., the profiles
flatten off. These so--called `shallow cusps' have been seen in {\it
HST} observations of several galaxies (e.g. Forbes \etal 1995; Lauer
\etal 1995; Carollo \etal 1996). If we fit the V band profile with the
dual power--law of Lauer \etal we derive an inner slope of --0.2. For
M$_V$ = --22.6, this value lies on the correlation of inner slope versus
absolute V magnitude (Forbes \etal 1995). Although the origin for this
correlation is still subject to debate, it is not the result of dust
obscuration but rather some fundamental scaling relation in galaxies. 
NGC 5846 obeys this scaling relation.

\subsection{Color Distribution}

The distribution of GC V--I color reveals a bimodal structure, with 
peaks around V--I = 0.96 and 1.17 (Fig. 4). 
A KS test confirms that the blue subpopulation is statistically
significant and a reduced $\chi^2$ goodness-of-fit indicates that two
Gaussians (with six free parameters) are a better fit to the color
distribution than one Gaussian (with three free parameters).  
Assuming the
color--metallicity relation of Couture \etal (1990), 
these colors correspond to metallicities of [Fe/H] = --1.2 and --0.2
respectively. 
In order to study the blue and
red subpopulations separately, we have defined a blue sample of 
232 GCs with colors of 0.84 $<$ V--I $\le$ 1.06 and a red sample of 361 with 
1.06 $<$ V--I $\le$ 1.28. 
The ratio of the red to blue samples changes from 1.70 
for the central pointing to 1.34 for the north and 1.54 for the 
south pointings. 
This tends to suggest that the red subpopulation is 
more centrally concentrated than the blue population. We measure a
small 
radial metallicity gradient, $\Delta$[Fe/H]/$\Delta$logR (arcsec) =
--0.13 $\pm$ 0.03. 
In the case of NGC 4472, a
metallicity gradient of --0.4 in $\Delta$[Fe/H]/$\Delta$logR (arcsec)  
was seen by Geisler \etal (1996) using Washington photometry. They
concluded that the gradient was probably due to the changing
proportions of the two subpopulations rather than a
smoothly varying radial gradient.

Next we examine whether or not the blue (metal--poor) and red (metal--rich)
samples have different luminosity functions. 
The dependence of GC luminosity on metallicity pertains to various aspects
of GC formation and evolution i.e. self--enrichment, the extent to which
massive GCs can retain heavy elements from the first generation of stars
and the ability of proto--globular 
clusters to sweep up enriched material during passage
near the galaxy center. 
There is, at present, no generally accepted model of GC formation
and no firm physical basis for assuming that the
GCLF is universal. It is generally assumed that the mass-to-light
ratio of all GCs is the same and, therefore, that the mass and
luminosity functions have the same form.  
For Milky Way and M31 GCs there is 
no luminosity--metallicity dependence 
(Huchra, Brodie \& Kent 1991), 
but this question remains largely unaddressed
for galaxies beyond the Local Group.
Bimodal color, and by inference metallicity, distributions have been
reported in a number of galaxies (see the list of Kissler--Patig 1996)
but a comparison of the luminosity
functions of the two subpopulations has so far been attempted only for
M87. Whitmore \etal (1995) and Elson \& Santiago (1996) each find evidence
for bimodal color distributions in their WFPC2 images of the M87 GC
system. 
They also claim that the red subpopulation is on average fainter 
than the blue subpopulation by 0.13 (Whitmore \etal ) and 0.3 (Elson
\& Santiago). 

In Fig. 10 we show the V band GCLF for the blue and red subpopulations
of NGC 5846 compared
to a Gaussian fit to the full sample in Paper I. Although there is
considerable scatter in each magnitude bin, both the blue and red
subpopulations appear to be consistent with the same Gaussian
(i.e. fixed turnover magnitude and dispersion). This is confirmed by a 
K--S test which indicates that the two subpopulations are {\it not} 
statistically
different. In order to test our ability to detect an offset in the
two GCLFs we have progressively faded the red sample. These tests
indicate that the two samples become statistically different 
only for offsets greater than 0.3 mag, i.e. we can not confirm offsets
of less than 0.3 magnitudes. Elson \&
Santiago (1996) do not quote the statistical significance of their 0.3
magnitude offset. For NGC 5846, we  
conclude that the blue and red GC populations 
have similar luminosity functions. There is no strong evidence for a
luminosity--metallicity relation for GCs within a single galaxy's
GC system. This does not, however, 
rule out the possibility of a GC luminosity--metallicity relation {\it
between} galaxies. Indeed such a relationship may allow us to understand 
the connection between GCLF turnover magnitude, GC metallicity
(Ashman, Conti \& Zepf 1995) and parent galaxy total luminosity
(Secker \& Harris 1993).

\subsection{Spatial Distribution}

The radial variations of GC properties may constrain
theories of GC formation and destruction.  
In Paper I we showed that the GCLF did not vary with 
galactocentric radius. Excluding the PC CCDs (which have a slightly brighter
detection limit), we now show in Fig. 8 that 
the mean GC magnitude is roughly constant, out to 30 kpc. If 
the mass-to-light ratio is 
constant, this implies that the mean mass of GCs 
does not vary with distance from the galaxy center. 
A similar 
result was found by Forbes \etal (1996) for a sample of 14 nearby 
ellipticals. 
Destruction 
processes, such as dynamical friction, appear to be 
insignificant over the radial range probed by our data. 
This result, and Forbes \etal (1996), appear to rule out the model of
Murali \& Weinberg (1996) on GC evolution. 

Fig. 5 shows that the metallicity (as derived from V--I colors) has a
slight dependence on galactocentric radius. 
This gradient is generally smaller than that measured in other giant
elliptical and cD galaxies, but these results seem to be consistent
with the observed trend of metallicity with specific frequency (see
Forbes, Grillmair \& Brodie 1996). 
Gradients are expected in both 
the Ashman \& Zepf (1992) merger and dissipational collapse pictures. 
However, projection effects tend to reduce such gradients (e.g. Secker
1996). 
The Searle \& Zinn 
(1978) picture of coalescing subunits, on the other hand, predicts no 
radial metallicity gradient.

The surface density of GCs in log space is shown in Fig. 6. A fit 
to all of the data points gives a slope of --0.7. 
If we exclude the inner data point, which lies at the radius
expected for the GC system core size (see Forbes \etal
1996), the slope steepens slightly to --0.8. 
Even at --0.8, the surface density slope is, to our knowledge, the flattest 
measured for GC systems 
(Kissler--Patig 1996). 
We have not made any correction for background galaxy contamination.  
We note that Haynes \& Giovanelli (1991) have studied the region
around the NGC 5846 group. They find both nearby and distant galaxies 
to the north and west of NGC 5846. Our data do not cover the west of
NGC 5846 but do include a northern pointing. Indeed there appears to
be an enhancement of faint galaxies in CCDs 2 and 3 in the north
pointing. 
Our selection process, 
after applying the size and color cuts, excludes the obvious resolved
galaxies. It is possible that compact background galaxies remain in
our object list in the northern pointing. However, the surface density
in the northern pointing is similar to that of the southern pointing
at a given galactocentric radius, including that background galaxy
contamination, if any, is negligibly small. 
The WFPC2 geometry makes it difficult to measure the galaxy
starlight slope, but we estimate it to be $\sim$ --1.5. 
Thus the GC system appears to have a much 
flatter surface density slope and is more extended
than the galaxy light. 
This situation is common for elliptical galaxies (e.g
Grillmair \etal 1994).  
Since the galaxy light
falls off faster with radius than the density of GCs, we would expect
the local S$_N$ value (i.e. the number of GCs normalized by the 
galaxy light out to that radius) to increase with radius. This is indeed
seen in Fig. 7, in which the local S$_N$ increases from $\sim$ 0.2 at
1 kpc to 2 at 12 kpc. We expect the S$_N$ value to increase smoothly
to the global value of 4.3 $\pm$ 1.1 at a radius of 50 kpc. 
The total number of GCs out to 50 kpc is estimated to be 4670 $\pm$
1270, 
which is
roughly twice the estimate of Harris \& van den
Bergh (1981) from photographic plates. 
We conclude that NGC 5846 has a much lower specific frequency than
other dominant ellipticals in rich clusters but is similar to dominant
ellipticals in groups (e.g. Bridges \& Hanes 1994).

The azimuthal distribution of GCs is generally consistent with a
spherical distribution, which is similar to the underlying galaxy 
light i.e., type E0--1. However,  there is 
evidence for a slight 
enhancement of counts around position angle 20$^{\circ}$ and a deficit
at 100$^{\circ}$. If this effect is real, the GCs appear to be 
aligned closer to the
galaxy's minor axis than its major axis (P.A. $\sim$ 100$^{\circ}$). 
The nearest galaxy to NGC 5846 is NGC 5846A, which lies to 
the south (P.A. $\sim$
190$^{\circ}$).

\subsection{Implications for Globular Cluster Formation}

One of main results from this study is that NGC 5846 has a bimodal
color and, by inference, metallicity distribution. A handful of other 
galaxies (usually having high luminosity and  high S$_N$) also reveal 
bimodality. This
indicates that there has been more than one epoch of GC formation in
these galaxies. A bimodal metallicity distribution is a 
prediction of the Ashman \& Zepf (1992) merger model. In this model
the blue GCs are thought to be the metal--poor population from the
progenitor galaxies and the red GCs are metal--rich ones formed in the
merging process. The red GCs are expected to be more centrally
concentrated than the blue GCs (which is hinted at in NGC 5846). 
However, this model has some difficulty in explaining other
features. For example, to produce a luminosity of M$_V$ = --22.6 the
merging of $\sim$ 10~L$^{\ast}$ spirals would be required. It is
difficult to imagine a clear bimodal GC distribution emerging from
such a history. The presence
of filamentary dust in the central regions suggests a recent encounter
with a gas--rich galaxy. The short dynamical times in this
region ($\le$ 10$^8$ yrs) and the fact that the GCs appear to be 
over 1 Gyr old
suggest that GCs did not form from this event. 

As mentioned in the introduction, M87 and NGC 5846 have a similar
X--ray luminosity and cooling flow rate and yet M87 has about three
times as many GCs per unit starlight as NGC 5846. These results
support the view of Bridges \etal (1996) that the bulk of GCs did {\it
not} form in a cooling flow. 
Other
mechanisms (such as a two--phase collapse or the accretion of external
GCs) should be considered to explain the GC bimodality in NGC
5846. We note that NGC 5846A, a compact E2 galaxy with M$_V$ $\sim$
--18.5, lies $\sim$ 6 kpc south of NGC 5846. The GCs of NGC 5846A may 
have contributed to those of NGC 5846 via tidal stripping and may
contribute even more if the whole
galaxy is eventually accreted. Unfortunately, NGC 5846A is not covered
by our WFPC2 images.

\section{Conclusions}

Using deep multi--filter WFPC2 images, we discuss the color and
spatial distributions of the globular cluster system in NGC 5846. 
We find a bimodal
color distribution with peaks at V--I = 0.96 and 1.17 
in all three WFPC2 pointings. These roughly correspond to metallicities of
[Fe/H] = --1.2 and --0.2 respectively. Such a bimodal distribution
suggests 
that at least formation episodes
have contributed to the GC system.  
The red (metal--rich) population may be 
more centrally concentrated than the blue (metal--poor)
population. The radial metallicity gradient is
$\Delta$[Fe/H]/$\Delta$logR (arcsec) = --0.13 $\pm$ 0.03. 
Both subpopulations have the same luminosity functions,
within an uncertainty of 0.3 mag. Thus we find no evidence for a
luminosity--metallicity relation within an individual galaxy's
globular cluster 
system. Previous work on the Milky Way and M31 reached a similar
conclusion (Huchra, Brodie \& Kent 1991). For a constant mass-to-light
ratio, 
the lack of a mass--metallicity relation for globular clusters, 
allows several 
formation models to be ruled out (ones in which a 
metallicity cooling criterion imprints a
characteristic mass).

The surface density of GCs falls off much more slowly with radius 
than the galaxy
light. The best--fit power law gives a slope of $\sim$ --0.75 which
appears to be 
the flattest globular cluster systems known. 
The globular clusters appear to be better aligned with the galaxy's
minor axis than its major axis, although this result is of marginal
significance. 
We find that the specific frequency 
increases steadily with radius in the inner regions and reaches a 
global value of 4.3 $\pm$ 1.1 at 50 kpc.  Thus NGC 5846 has a much lower
specific frequency than other dominant ellipticals in clusters but is
similar to dominant ellipticals in groups. 

The galaxy isophotes deviate from pure ellipses and 
become slightly boxy (extra light at 45$^{\circ}$ to the major axis) 
in the central regions,
perhaps due to the filamentary dust. The dust is probably the result
of a past merger or accretion of a gas--rich galaxy and provides the
fuel for the compact radio core. 
We have searched for
proto--globular clusters associated with the dust (and gas) and find 
none. 

\noindent
{\bf Acknowledgments}\\
We thank C. Grillmair, M. Kissler--Patig and M. Rabban 
for helpful discussions. We also thank the referee, J. Secker, for his
comments. This
research was funded by the HST grant GO-05920.01-94A\\

\newpage
\noindent{\bf References}

\noindent
%Ajhar, E. A., Blakeslee, J. P., \& Tonry, J. L. 1994, AJ, 108, 2087 (ABT94)\\
%Ashman, K. M., \& Bird, C. M. 1993, AJ, 106, 2281\\
Ashman, K. M., \& Zepf, S. E. 1992, ApJ, 384, 50\\
%Aguliar, L., Hut, P., \& Ostriker, J. P. 1988, ApJ, 335, 720\\
Ashman, K. M., Conti, A., \& Zepf, S. E. 1995, AJ, 110, 1164\\ 
%Baum, W. A., \etal 1995, AJ, 110, 2537\\
Bender, R., Burstein, D., \& Faber, S. M. 1992, ApJ, 399, 462\\
Biermann, P. L., Kronberg, P. P., \& Schmutzler, T. 1989, A \& A, 208,
22\\
%Binggeli, B, Tammann, G. A., \& Sandage, A. 1987, AJ, 94, 251\\
%Bender, R. 1988, A \& A, 202, L5\\
%Bender, R. 1990, Dynamics and Interactions of Galaxies, p. 232, ed. 
%R. Wielen, Springer-Verlag, Berlin\\
%Bender, R., Dobereiner, S., \& Mollenhof, C. 1988, A \& AS, 74, 385\\ 
%Brodie, J. P., \& Huchra, J. 1991, ApJ, 379, 157\\
Bridges, T. J., \& Hanes, D. A. 1994, ApJ, 431, 625\\
Bridges, T. J., Hanes, D. A., \& Harris, W. E. 1991, AJ, 101, 469\\
Bridges, T. J., Carter, D., Harris, W. E., \& Pritchet, C. J. 1996,
MNRAS, in press\\
Bruzual, G., \& Charlot, S. 1996, in preparation\\
%Burrows, C., \etal 1993, Hubble Space Telescope Wide Field and
%Planetary Camera 2 Instrument Handbook, STScI\\
Carollo, C. M., Franx, M., Illingworth, G. D., \& Forbes, D. A. 1996,
ApJ, submitted\\
Couture, J., Harris, W. E., \& Allwright, J. W. B., 1990, ApJS, 73,
671\\
%Couture, J., Harris, W. E., \& Allwright, J. W. B., 1991, ApJ, 372, 97\\
%Davies, R. L., \etal 1987, ApJS, 64, 581\\
%Carollo, C. M., Danziger, I. J., \& Buson, L. 1993, MNRAS, 265, 553\\
Elson, R. A. W., \& Santiago, B. X. 1996, MNRAS, 280, 971\\
%Forbes, D. A. 1991, MNRAS, 249, 779\\
Faber, S. M., \etal 1989, ApJS, 69, 763\\
Fabian, A. C., Nulsen, P. E. J., \& Canizares, C. R. 1984, Nature,
310, 733\\ 
Fisher, D., Franx, M., \& Illingworth, G. D. 1995, ApJ, 448, 119\\
%Forbes, D. A. 1994, AJ, 107, 2017\\
Forbes, D. A., \& Thomson, R. C. 1992, MNRAS, 254, 723\\
%Forbes, D. A., Reitzel, D. B., \& Williger, G. M. 1994, AJ, in press\\
Forbes, D. A., Franx, M., \& Illingworth, G. D. 1995, AJ, 109, 1988\\
%Forbes, D. A. 1996a, AJ, 112, 954\\
%Forbes, D. A. 1996b, AJ, in press\\
Forbes, D. A., Brodie, J. P., \& Huchra, J. 1996, AJ in press\\
Forbes, D. A., Grillmair, C. J.,  \& Brodie, J. P. 1996, in preparation\\
%Forbes, D. A., Elson, R. A. W., Phillips, A. C., 
%Illingworth, G. D. \& Koo, D. C. 1994, ApJ, 437, L17\\
Forbes, D. A., Franx, M., Illingworth, G. D., \& Carollo, C. M. 1996,
ApJ, in press\\
%Forbes, D. A., Sparks, W. B., \& Macchetto, F. D. 1990, Paired and
%Interacting Galaxies, p. 431, ed. J. W. Sulentic, W. C. Keel and C.
%M. Telesco, NASA conference publication 3098\\
Forte, J. C., Martinez, R. E., \& Muzzio, J. C. 1982, AJ, 87, 1465\\
%Franx, M., \& Illingworth, G. D. 1988, ApJ, 327, L55\\
%Fleming, D. E. B., Harris, W. E., Pritchet, C. J., \& Hanes,
%D. A. 1995, AJ, 109, 1044\\
%Freeman, W. L. \etal 1994, ApJ, 427, 628\\ 
%Goudfrooij, P., Norgaard Nielson, H. U., Hansen, L., Jorgensen, H. E.,
%\& de Jong, T. 1990, A \& A, 228, L9\\
%Goudfrooij, P., Hansen, L., Jorgensen, H. E., \& Norgaard Nielson, H. U.  
%1994, A \& AS, 105, 341\\
Goudfrooij, P., \& Trinchieri, G. 1996, in preparation\\
%Gunn, J. E. 1979, Active Galactic Nuclei, p. 213, ed.\ C.\ Hazard and
%S.\ Mitton, Cambridge University Press, Cambridge\\
%Grillmair, C., Pritchet, C., \& van den Bergh, S. 1986, AJ 91, 1328\\
Geisler, D., Lee, M. G., \& Kim, E. 1996, AJ, in press\\
Grillmair, C. \etal 1994, AJ, 108, 102\\ 
%Grillmair, C. \etal 1994b, ApJ, 422, L7\\ 
%Grillmair, C. J. 1995, personal communication\\
%Guzm\'an, R., Lucey, J. R., \& Bower, R. G. 1993, MNRAS, 265, 731\\
%Hanes, D. A. 1977, Mem. RAS, 84, 45\\
%Harris, H. C., Baum, W. A., Hunter, D. A., \& Kreidel, T. J. 1991, AJ,
%101, 677\\ 
%Harris, W. E. 1990, PASP, 102, 966\\
%Harris, W. E. 1991, ARAA, 29, 543\\
%Harris, W. E. 1993, The Globular Cluster -- Galaxy Connection, p. 472,
%ed. G. Smith and J. Brodie, ASP conference series, San Francisco\\
%Harris, W. E. \etal 1986, AJ, 91, 822\\ 
%Harris, W. E., \& Pudritz, R. E. 1994, 429, 177\\ 
%Harris, W. E., Allwright, J. W. B., Pritchet, C. J., \& 
%van den Bergh, S. 1991, ApJS, 276, 491\\
%Huchra, J., Davis, M., Latham, D., \& Tonry, J. 1983, ApJS, 52, 89\\
%Harris, W. E., Pritchet, C. J., \& McClure, R. D. 1995, ApJ, 441, 120\\
Harris, W. E., \& van den Bergh, S. 1981, AJ, 86, 1627\\
%Hau, G. K. T., \& Thomson, R. C. 1994, MNRAS, 270, L23\\
Haynes, M. P., \& Giovanelli, R. 1991, AJ, 102, 841\\
Holtzman, J., \etal 1992, AJ, 103, 691\\
Holtzman, J., \etal 1996, AJ, 112, 416\\
%Holtzman, J., \etal 1995a, PASP, in press\\
%Holtzman, J., \etal 1995b, PASP, submitted\\
Huchra, J., Brodie, J. P., \& Kent, S. 1991, ApJ, 370, 495\\
%Illingworth, G. D., \& Franx, M. 1989, Dynamics of Dense Stellar
%Systems, p. 13, ed. D. Merritt, Cambridge University Press,
%Cambridge\\
%Jacoby, G. H., \etal 1992, PASP, 104, 599 (J92)\\
%Kelson, D. D. \etal 1996, ApJ, 463, 26\\
Kissler--Patig, M., 1996, A \& A, in press\\
Lauer, T., \etal 1995, AJ, 110, 2622\\
Longo, G., \& de Vaucouleurs, A. 1983, A General Catalogue of
Photometric Magnitudes and Colors in the UBV system (University of
Texas, Austin)\\
%Mould, J. R. \etal 1995, ApJ, 449, 413\\
Mollenhoff, C., Hummel, E., \& Bender, R. 1992, A \& A, 255, 35\\
McLaughlin, D. E., Harris, W. E., \& Hanes, D. A. 1993, ApJ, 409, L45\\
McLaughlin, D. E., Secker, J., Harris, W. E., \& Geisler, D. 1995, AJ,
109, 1033\\
Murali, C., \& Weinberg, M. D. 1996, MNRAS, in press\\
%Murray, S. D., \& Lin, D. N. C. 1992, ApJ, 400, 265\\
%Perelmuter, J. L. 1995, ApJ, 454, 762\\
%Perelmuter, J. L., Brodie, J. P., \& J. P. 1995, AJ, 110, 620\\
%Sandage, A., \& Tammann, G. A. 1995, ApJ, 446, 1\\
Searle, L., \& Zinn, R. 1978, ApJ, 437, 214\\
Secker, J. 1996, ApJL, 469, 81\\
%Secker, J. 1992, AJ, 104, 1472\\
Secker, J., \& Harris, W. E. 1993, AJ, 105, 1358 (SH93)\\
%Schweizer, F. 1987, Nearly Normal Galaxies, p. 18, ed. S. Faber,
%Springer-Verlag, New York\\ 
Stetson, P. B., 1987, PASP, 99, 191\\
%Surma, P. 1992, Structure, Dynamics and Chemical Evolution of
%Elliptical Galaxies, ed. I. J. Danziger, W. W. Zeilinger and K. Kjar,
%ESO: Garching, p. 669\\
%Tonry, J. L., Ajhar, E. A., \& Luppino, G. A. 1990, AJ, 100, 1416\\
%Tonry, J. L. 1991, ApJ, 373, L1\\
%van den Bergh, S. 1995, AJ, 110, 2700\\
West, M. J., Cote, P., Jones, C., Forman, W., \& Marzke, R. O. 1995,
ApJ, 453, L77\\
Whitmore, B. C., Schweizer, F., Leitherer, C. Borne, K., \& Robert,
C. 1993, AJ, 106, 1354\\
Whitmore, B. C., \& Schweizer, F. 1995, AJ, 109, 960\\
Whitmore, B. C., Sparks, W. B., Lucas, R. A., Macchetto, F. D., \&
Biretta, J. A. 1995, ApJ, 454, L73\\
%Zepf, S. E., \& Ashman, K. M. 1993, MNRAS, 264, 611\\
Zepf, S. E., Ashman, K. M., \& Geisler, D. 1995, ApJ, 443, 570\\

%\noindent{\bf Figure Captions}

\begin{figure*}[p]
\centerline{\psfig{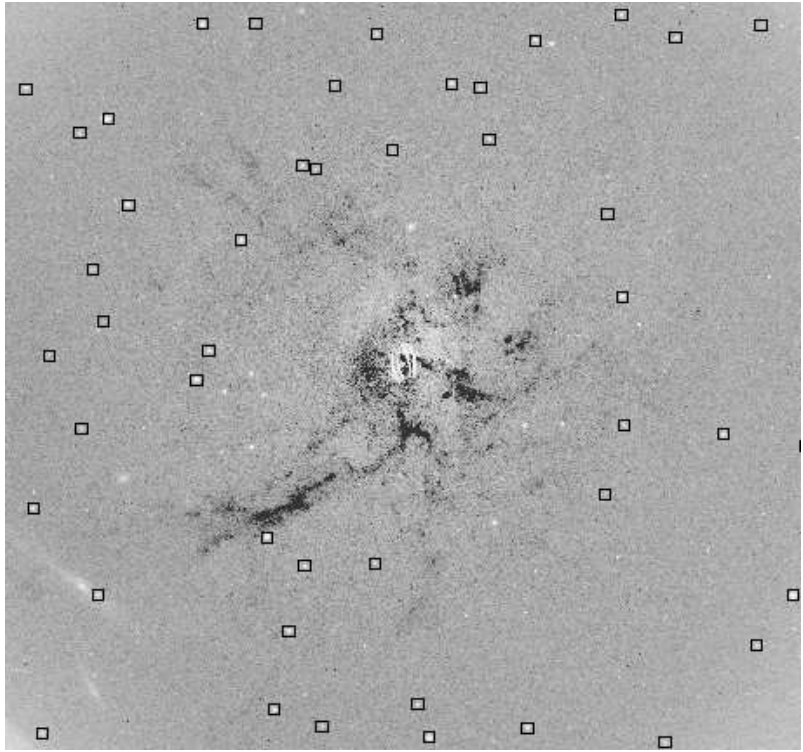}}
\caption{\label{fig1}
Grey scale WFPC2 image of the PC CCD (about 30$^{''}$ $\times$
30$^{''}$ area) of the central region of NGC 5846. The automatically 
selected globular clusters beyond 7$^{''}$ (1 kpc) from the galaxy
center are indicated by black squares. Several radial dust filaments
can be seen reaching into the nucleus. 
}
\end{figure*}

\begin{figure*}[p]
\centerline{\psfig{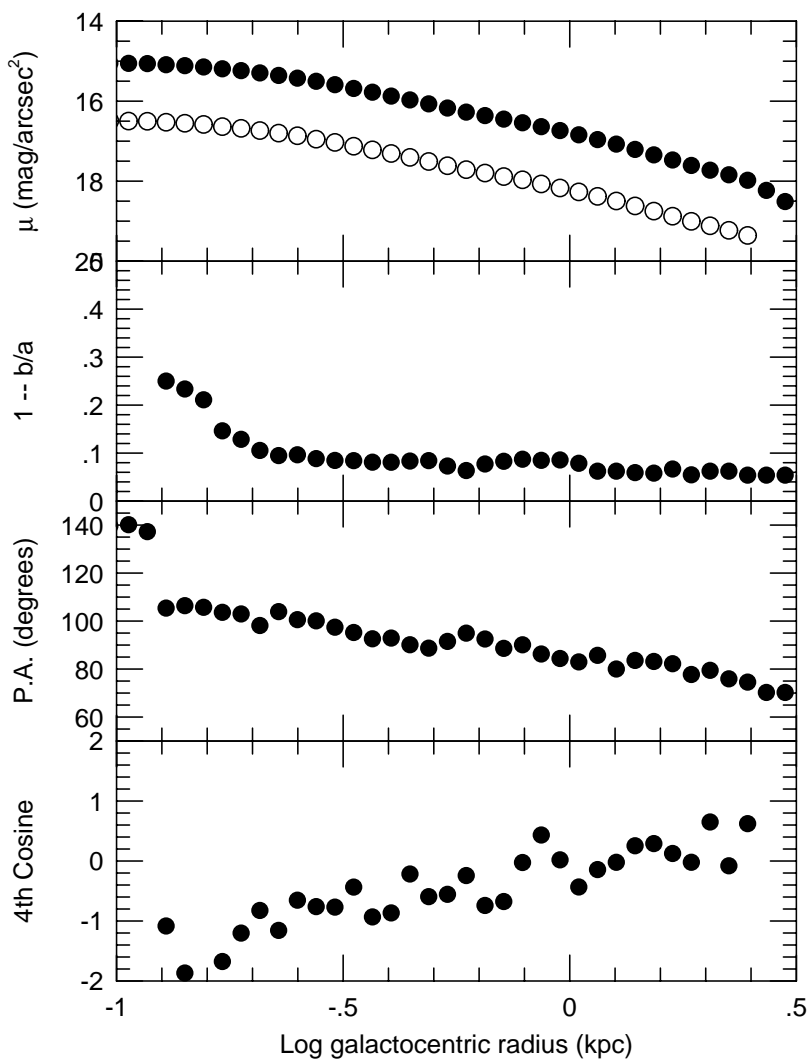}}
\caption{\label{fig2}
Galaxy properties within 20$^{''}$ (3 kpc). We show surface
brightness, ellipticity, position angle and the 4th cosine term
deviations from a pure ellipse. 
The I band data are
represented by filled circles and the V band surface brightness data
by open circles. In the central regions, NGC 5846 has near constant
V--I color and ellipticity. The position angle changes from
70$^{\circ}$ to $\sim$ 100$^{\circ}$ near the nucleus, 
and 4th cosine term becomes more boxy. 
}
\end{figure*}

\begin{figure*}[p]
\centerline{\psfig{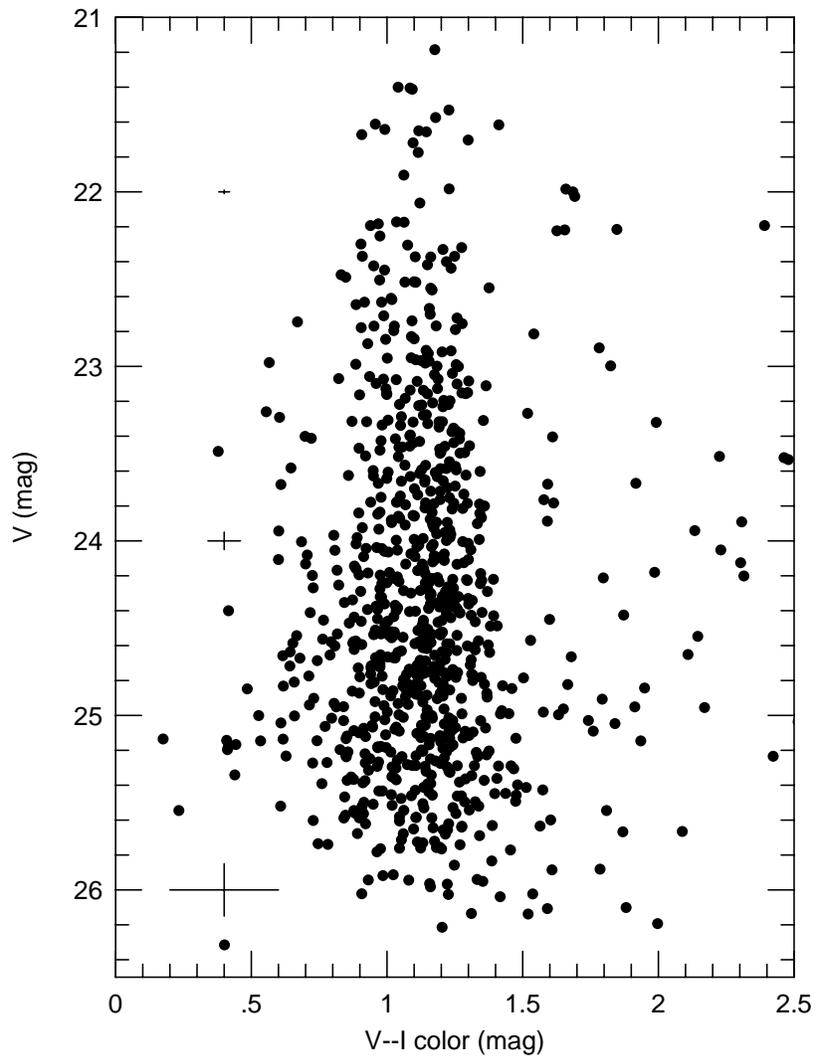}}
\caption{\label{fig3}
Color--magnitude diagram for a sample of 837 globular clusters around NGC
5846. Typical error bars are shown on the left. The globular cluster
color is roughly constant with magnitude. 
}
\end{figure*}

\begin{figure*}[p]
\centerline{\psfig{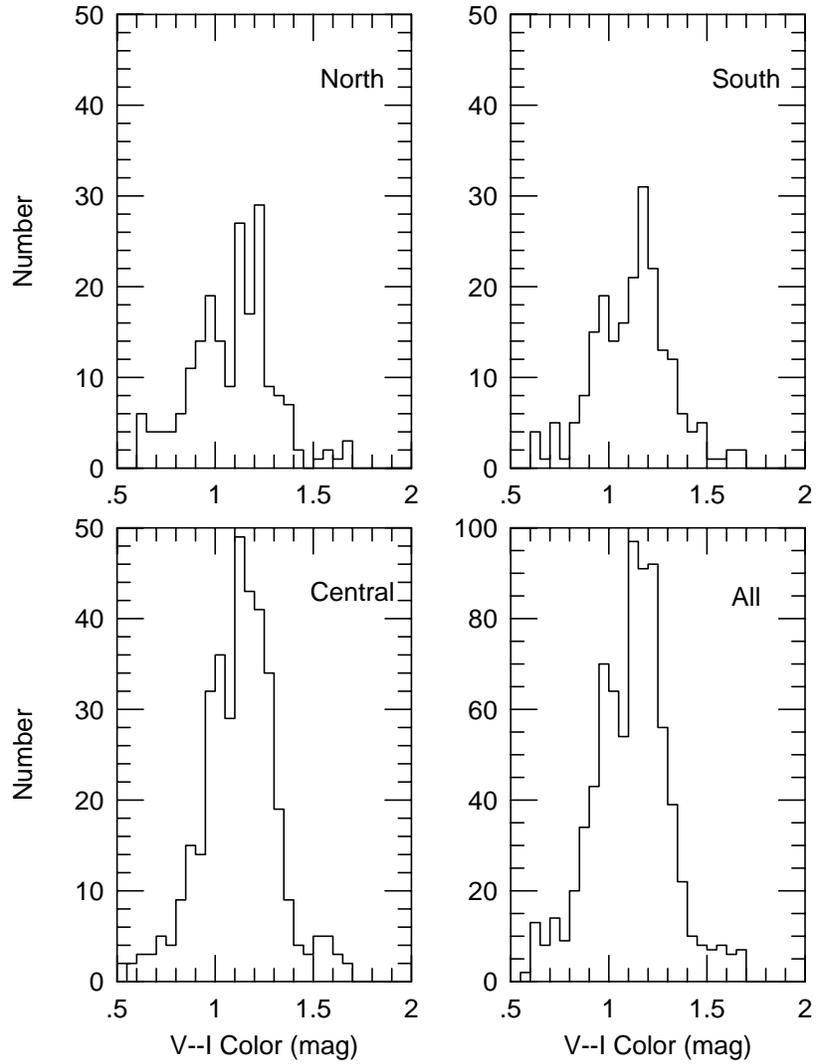}}
\caption{\label{fig4}
Histograms of globular cluster V--I colors (within 3$\sigma$ of the
mean color). The four panels show data from the north, south and
central pointings, along with the combined sample. A clear bimodal
color distribution can be seen.  
}
\end{figure*}

\begin{figure*}[p]
\centerline{\psfig{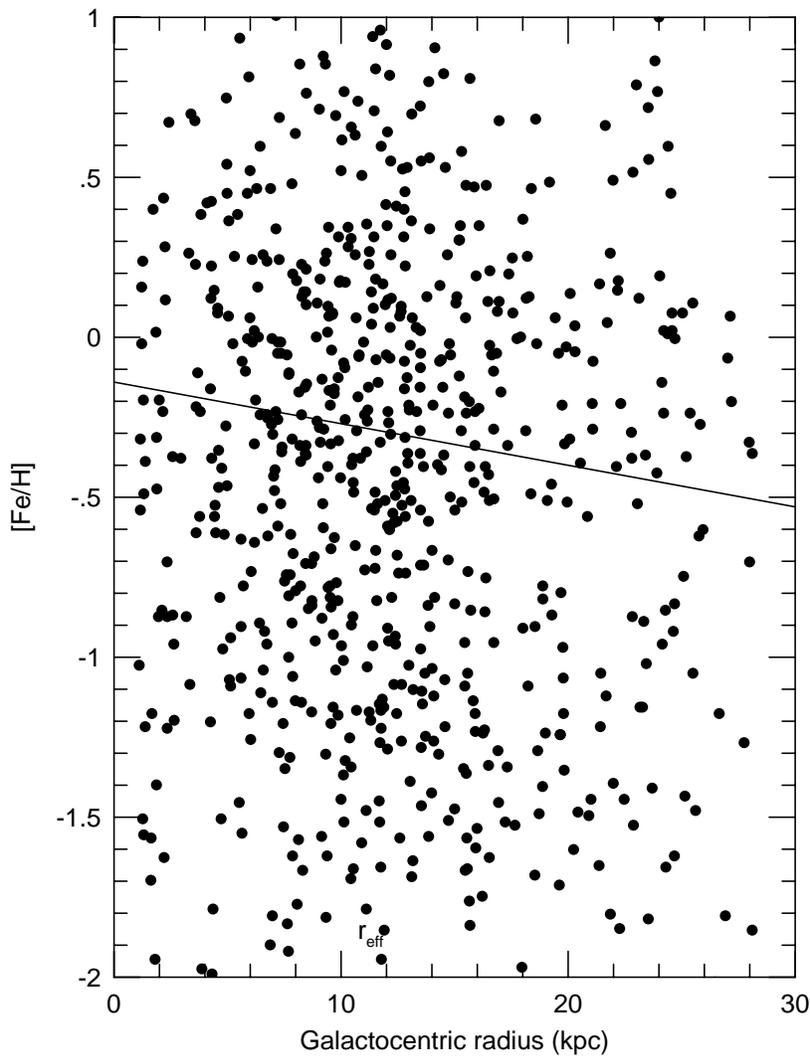}}
\caption{\label{fig5}
Radial variation of globular cluster metallicity. 
The solid line represents a weighted fit to the data. There is a
shallow metallicity gradient. 
The effective radius of the galaxy (r$_{eff}$) is
also indicated. The typical random error of each data point is $\pm$
0.25. 
}
\end{figure*}

\begin{figure*}[p]
\centerline{\psfig{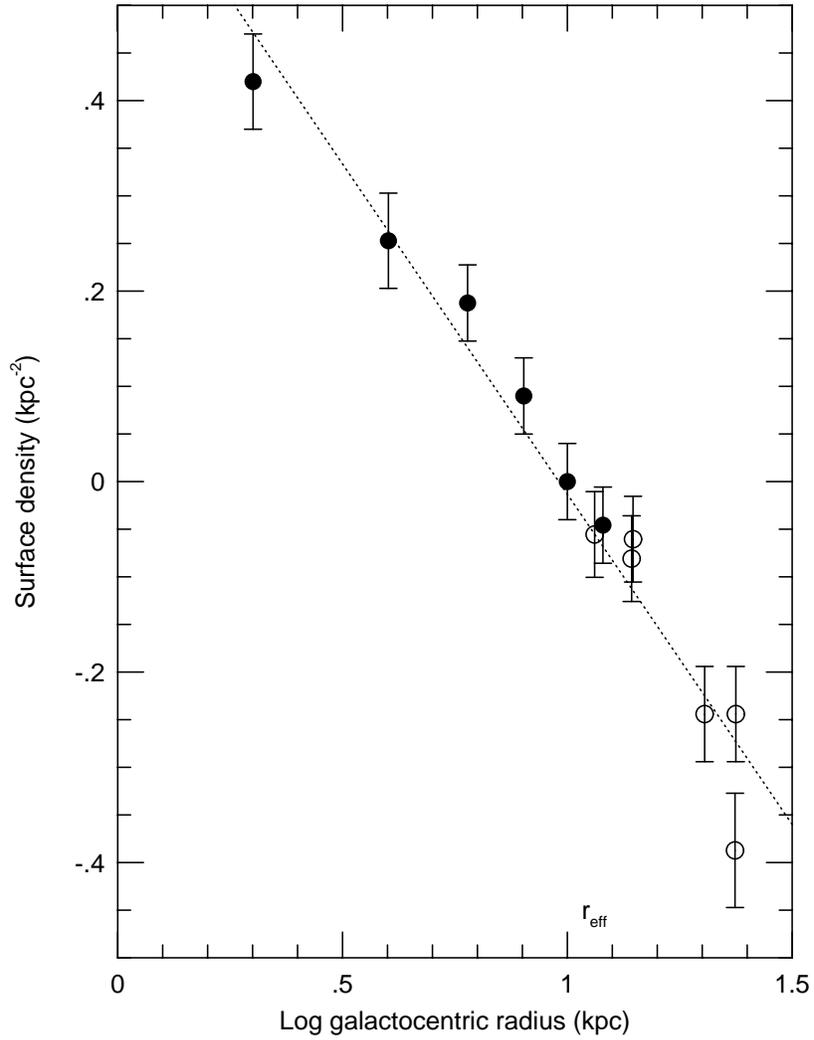}}
\caption{\label{fig6}
Surface density profile of globular clusters. The surface density has
been calculated using two different methods (see text for details)
which are represented by open and filled circles. Error bars represent
Poisson statistics. The dashed line is a best fit to all data points,
which gives a slope of --0.7. 
}
\end{figure*}

\begin{figure*}[p]
\centerline{\psfig{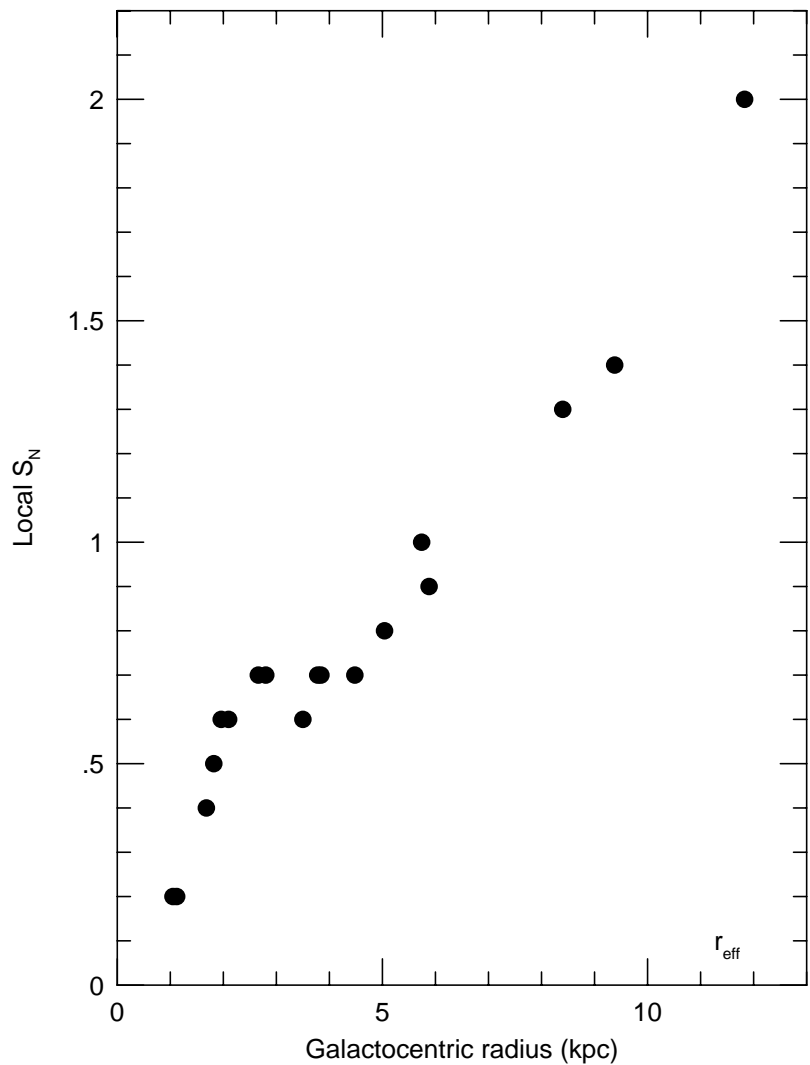}}
\caption{\label{fig7}
Radial variation of local specific frequency. The local specific frequency
is the number of globular clusters normalized by the integrated galaxy
light at each radius. The local S$_N$ increases with distance from the
galaxy center, reaching a global value of 4.3 at 50 kpc. 
}
\end{figure*}

\begin{figure*}[p]
\centerline{\psfig{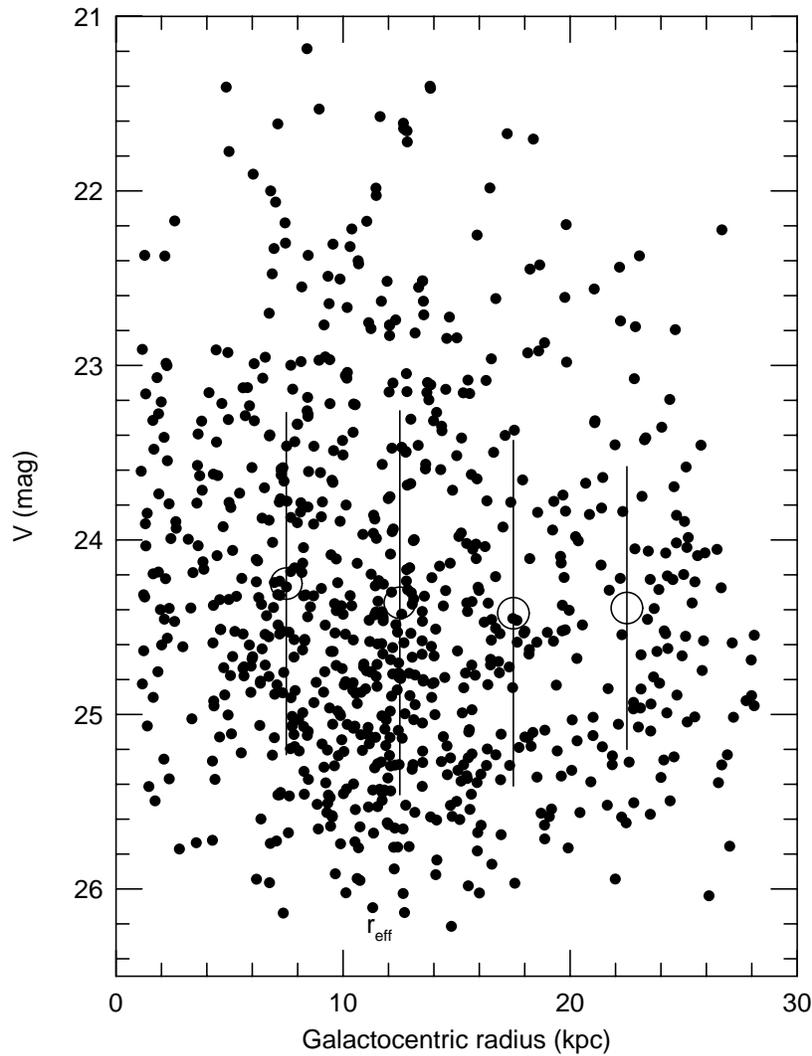}}
\caption{\label{fig8}
Radial variation of V magnitude. The mean magnitude in 4 bins is
indicated along with the rms scatter. 
There is little or no magnitude
gradient beyond the central PC CCD ($\sim$ 2.5 kpc), which has a
slightly brighter  detection limit. 
}
\end{figure*}

\begin{figure*}[p]
\centerline{\psfig{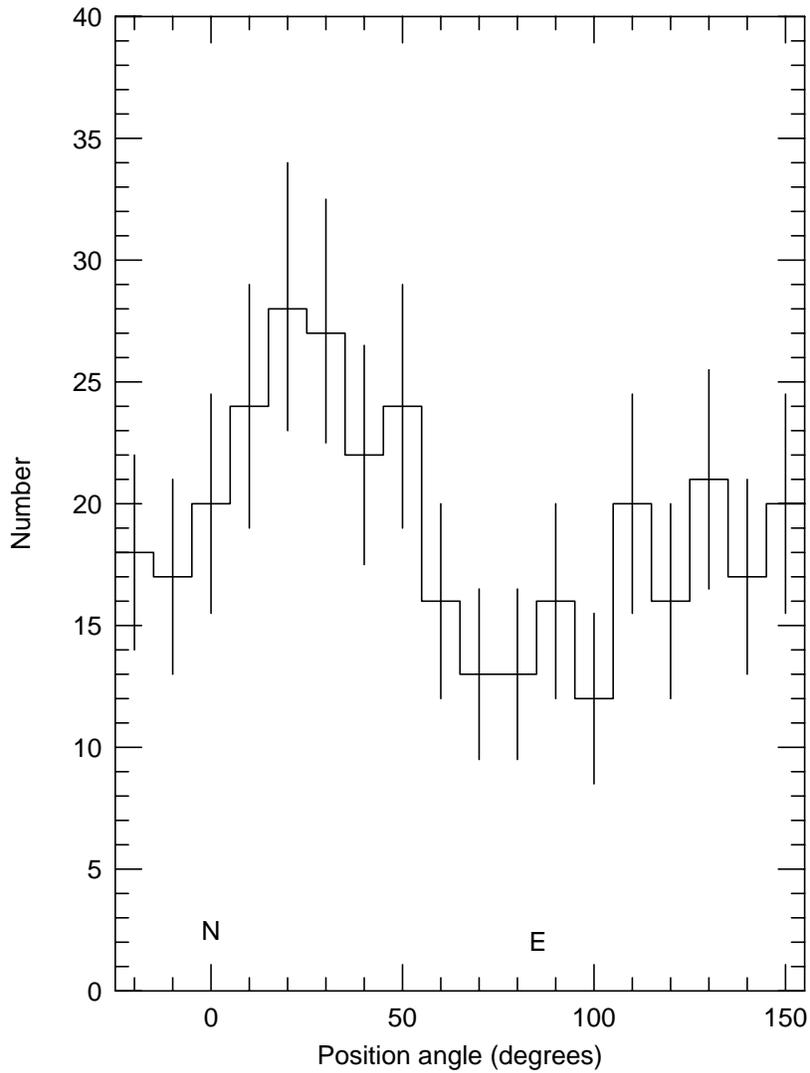}}
\caption{\label{fig9}
Histogram of globular cluster position angle. Error bars represent
Poisson statistics. North and East are indicated. 
The galaxy major axis lies at $\sim$ 100$^{''}$. 
The globular
clusters reveal a slight enhancement at $\sim$ 20$^{\circ}$ but
otherwise appear consistent with a spherical distribution. 
}
\end{figure*}

\begin{figure*}[p]
\centerline{\psfig{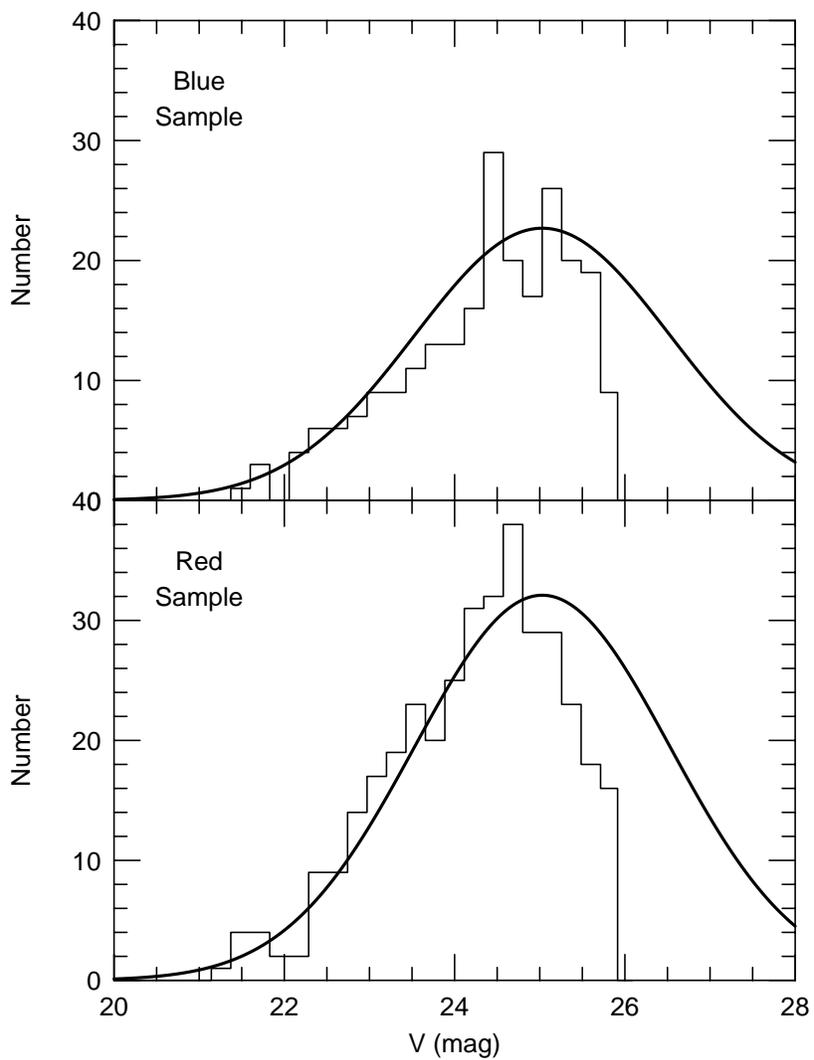}}
\caption{\label{fig10}
Globular cluster luminosity functions for the red and blue samples. 
A Gaussian profile with a turnover magnitude appropriate for the full
sample has been superposed and arbitrarily scaled  in the vertical
direction. 
}
\end{figure*}

%\begin{figure}
%\centerline{\psfig{figure=table1.epsi,width=300pt}}
%\caption{\label{}
%}
%\end{figure}

\end{document}